\documentclass[aps,showpacs,nofootinbib,prd,twocolumn,,showkeys]{revtex4-1}
\pdfoutput=1
\usepackage{graphicx}
\usepackage[retainorgcmds]{IEEEtrantools}
\usepackage{amssymb}
\usepackage{amsmath}
\usepackage[svgnames]{xcolor}
\usepackage{mathtools,slashed}
\usepackage{epstopdf}
\usepackage[utf8]{inputenc}
\usepackage{url}
\usepackage[colorlinks,citecolor=DarkGreen,linkcolor=DarkRed,urlcolor=DarkBlue]{hyperref}
\usepackage{subfiles}
\usepackage{physics}
\usepackage{tikz}
\usepackage[compat=1.1.0]{tikz-feynhand}
\usetikzlibrary{arrows.meta}
\usepackage{pgfplots}
\usepackage{xparse}

\usetikzlibrary{positioning,arrows,patterns}
\usetikzlibrary{decorations.markings}
\usetikzlibrary{calc}
\usetikzlibrary {arrows.meta}

\usepackage{newtxtext}
\usepackage{newtxmath}
\usepackage[normalem]{ulem}

\usepackage{epsfig}

\newcommand{\be}{\begin{eqnarray}}
\newcommand{\ee}{\end{eqnarray}}
\begin{document}

\title{Thermal and baryon density modifications to the $\sigma$-boson propagator: A road to describe the transfer of vorticity to spin in a nuclear environment in relativistic heavy-ion collisions
}

\author{Alejandro Ayala$^1$}
\author{Jos\'e Jorge Medina-Serna$^1$}
\author{Isabel Dom\'\i nguez$^2$}
\author{Ivonne Maldonado$^3$}
\author{Mar\'\i a Elena Tejeda-Yeomans$^4$}
\address{
$^1$Instituto de Ciencias
Nucleares, Universidad Nacional Aut\'onoma de M\'exico, Apartado
Postal 70-543, CdMx 04510,
Mexico}
\address{$^2$Facultad de Ciencias F\'\i sico-Matem\'aticas, Universidad Aut\'onoma de Sinaloa, Avenida de las
Am\'ericas y Boulevard Universitarios, Ciudad Universitaria, Culiac\'an, 80000, Mexico}
\address{$^3$Joint Institute for Nuclear Research, Dubna, 141980, Russia}
\address{$^4$Facultad de Ciencias-CUICBAS, Universidad de Colima, Bernal D\'\i az del Castillo No. 340,
Colonia Villas San Sebasti\'an, Colima, 28045, Mexico}

\begin{abstract}
In the context of the description of how the vortical motion, produced in peripheral heavy-ion collisions, is transferred to the spin of hadrons, we compute the $\sigma$-meson propagator at finite temperature and baryon density. This propagator encodes the properties of a medium consisting mainly of nucleons{, and can be used to model the main interactions between hadrons} in the corona region of the reaction. We compute the one-loop $\sigma$ self-energy in an approximation that accounts for the large nucleon mass. From the real part of the self-energy, we find the dispersion relation and show that the $\sigma$-mass receives a non-negligible thermal and baryon chemical dependent {contribution}. From the imaginary part, we also compute the spectral density, which we show to contain a piece coming from the branch cut associated with Landau damping. We also present approximations for the dispersion relation and the residue at the pole in the small- and large-momentum regimes and complement the calculation, providing the sum rules satisfied by the propagator. This study aims to determine one of the elements needed to compute how the vortical motion in the corona region of the reaction is transferred to the spin of $\Lambda$ hyperons that can interact with nucleons by $\sigma$-meson exchange.
\end{abstract}


\maketitle

\section{Introduction}

The possibility that an intense vortical motion is generated in the interaction region of semi-central heavy-ion collisions has been  {examined extensively~\cite{Becattini2008,Becattini:2015ska,Becattini:2014yxa,Becattini2017,Huang:2020xyr,Sass:2022ucj,Karpenko:2021wdm}. In these reactions, the different profiles of matter density on the transverse plane of the colliding nuclei induces the development of a global angular momentum. This, in turn, can be transferred to the spin of particles created in the reaction, producing a net polarization along the transverse plane. The well-known Barnett
effect, by which a spinning ferromagnet experiences a change in its magnetization~\cite{1915PhRv....6..239B} and the closely related Einstein–de Haas effect, where a change in the magnetic moment of a free body causes the body to rotate~\cite{1915KNAB...18..696E}, support this expectation. The study of polarization phenomena in relativistic heavy-ion collisions has opened a new avenue of research that promises quantify better the properties of the matter created in these reactions~\cite{Becattini:2024uha,Niida:2024ntm}. }

An excellent probe to measure the transfer of vorticity to the spin of hadrons is provided by the $\Lambda$ (or $\Bar{\Lambda}$) hyperon. This particle is special since, for its main decay channel $\Lambda\rightarrow p+\pi^-$, the proton tends to be emitted along the direction of the spin of the $\Lambda$. The measurements of a small but finite $\Lambda$ and $\Bar{\Lambda}$ global polarization, which increase as the collision energy decreases, were first reported in Ref.~\cite{STAR:2017ckg}. Since then, several other experiments have found a nonzero global polarization that shows a similar increasing trend as the collision energy decreases~\cite{CMS:2025nqr,Gou:2024yrx,HADES:2022enx,STAR:2021beb,ALICE:2019onw,STAR:2018gyt}. Furthermore, it has been recently shown that the Multi-Purpose Detector (MPD)~\cite{MPD:2022qhn}, soon to enter into operation at the Nuclotron-based Ion Collider Facility (NICA), can reconstruct the global $\Lambda$ polarization and consequently, a handful of estimations for this observable have been performed in the NICA energy  range~\cite{Tsegelnik:2024ruh,Troshin:2024nig,Nazarova:2024jic,Nazarova:2021lry,Ayala:2020vvs}.

The experimental results have motivated the search for explanations of the origin of the generated angular momentum and of the $\Lambda$ polarization, making use of several of approaches that can be divided in two camps: the Monte Carlo and the phenomenological venues. In the former, {several generators have been used}, among them, the Multiphase Transport model (AMPT) sometimes combined with a hydro evolution to compute the global hyperon polarization in the collision energy range $\sqrt{s_{NN}}=1$ GeV -- 2.76 TeV.~\cite{Guo:2022cxa,Wu:2020yiz,Wei:2018zfb,Xia:2018tes}. Other calculations of vorticity and global polarization for energies between $\sqrt{s_{NN}}=$3 GeV - 200 GeV use the Ultra-relativistic Quantum Molecular Dynamics (UrQMD) model combined with viscous hydro~\cite{Deng:2021miw,Deng:2020ygd,Karpenko:2016jyx}. The Heavy Ion Jet Interaction Generator (HIJING) and the microscopic transport model PACIAE~\cite{Lei:2021mvp,Deng:2016gyh} have also been used. {In the latter, attention has been paid mainly to the origin of the mechanism responsible for the transfer of rotational motion to spin,  which can only happen provided that the reaction induced by the medium occurs fast enough such that the alignment of the spin and the angular velocity takes place within the lifetime of the medium~\cite{Montenegro:2018bcf,Kapusta:2019ktm,Kapusta:2019sad,Montenegro:2020paq,Kapusta:2020dco,Kapusta:2020npk,Torrieri:2022ogj}.}


Despite the large activity on the subject, the microscopic mechanism that produces the polarization from vorticity is not fully understood. Furthermore, the seemingly different polarizations measured for $\Lambda$ and $\Bar{\Lambda}$ could signal that different production mechanisms influence the final polarization. Since by the time hadronization takes place, the strength of a possible magnetic field produced in the reaction is small, it is unlikely that this difference is due to the presence of such a field. In a series of recent works~\cite{Ayala:2023xyn,Ayala:2021xrn,Ayala:2020soy}, we have put forward the idea that a two-component interaction region, consisting of a high-density core and a dilute corona that acts as the source of $\Lambda$s and $\Bar{\Lambda}$s, could be behind the different polarization strengths. In this core-corona model, the high-density core {consists} of deconfined matter, whereas the low-density corona consists mainly of nucleons. Both regions participate in the vortical motion. 

$\Lambda$s and $\Bar{\Lambda}$s coming from the core are assumed to be formed by the coalescence of a $s$-quark with a $u d$-diquark and to 
maintain the direction of the spin provided by the $s$-quark. In this scenario, it is possible to resort to a field-theoretical calculation to compute the relaxation time that an $s$-quark takes to align its spin with the direction of the angular momentum in a rotating medium~\cite{Ayala:2021osy} at finite temperature and baryon density. From this relaxation time, one can compute the intrinsic polarization of $\Lambda$s and $\Bar{\Lambda}$s coming from the core~\cite{Ayala:2023vgv,Ayala:2020ndx,Ayala:2019iin}, which can later on be used in the model to compute the global polarization.

The model assumes that hyperon polarization in the corona region is negligible. However, for the smallest measured collision energies, the model shows a discrepancy with data as a function of centrality, which we have attributed to the sudden drop polarization in the corona~\cite{Ayala:2023xyn}. Therefore, a better estimate of the way the vortical motion is transferred to the spin of $\Lambda$s and $\Bar{\Lambda}$s produced in the corona is required. In this work, we start to undertake this task. Since the interaction between nucleons and $\Lambda$s can be described as mediated by a $\sigma$-meson, and this in turn can change its properties when immersed in a high-density/temperature environment, we hereby compute the modification of the $\sigma$-meson propagator considering temperature and baryon density contributions. Effective models to describe the interaction of nucleons with $\sigma$-mesons have been extensively considered in the literature, see for example Refs.~\cite{Csernai:2018yok,Santos:2018mxl,Sun:2017fnf,Oertel:2014qza,Wang_2013,Song:2010qhk,Barros:2000iy}.

The work is organized as follows: In Se.~\ref{II}, we compute the one-loop $\sigma$-meson propagator at finite temperature and baryon density. We accomplish this by computing the self-energy of the in-medium $\sigma$ meson. From the real and imaginary parts of this self-energy, we compute the {temperature and baryon chemical potential dependent} dispersion relation and spectral density. In Se.~\ref{III} we study the analytical properties of the in-medium $\sigma$-meson propagator. In Se.~\ref{IV} we conclude and provide an outlook of the next-step use of this propagator for the computation of the angular momentum transferring to the $\Lambda$, $\Bar{\Lambda}$ spin in the context of relativistic heavy-ion collisions.

\section{One-loop thermal and baryon density corrections to the $\sigma$-meson}\label{II}

In the linear Relativistic Mean Field model, the Lagrangian density that describes the coupling of nucleons and $\sigma$ is
\begin{eqnarray}
    \mathcal{L}\!=\! \Bar{\psi}\left[i\gamma^\mu\partial_\mu-M_N-g_\sigma \sigma\right]\psi+\frac12 \partial_\mu\sigma\partial^\mu\sigma-\frac12 m_\sigma^2 \sigma^2,
\end{eqnarray}
where $\psi$ and $\sigma$ are the nucleon and the $\sigma$-meson fields, respectively, $M_N$ and $m_\sigma$ are the nucleon and $\sigma$-mass, respectively, and $g_\sigma$ is the coupling between the nucleon and the $\sigma$-meson. Considering the corona as a medium where nucleons and $\sigma$s interact at finite temperature $T$ and baryon chemical potential $\mu$, the one-loop effective $\sigma$ propagator $\Delta^*$ can be written, in the imaginary time formalism of finite temperature field theory, as
\begin{eqnarray}
\Delta^*(i\omega,p)=\frac{1}{\omega^2+p^2+m_\sigma^2+\Pi},
\label{effectiveprop}
\end{eqnarray}
where $\Pi$ is the $\sigma$ self-energy, which at one-loop order, depicted in Fig~\ref{sigma_Selfenergy}, is given by  
\begin{eqnarray}
\Pi&=&-g_\sigma^2\sum_n\int\frac{d^3k}{(2\pi)^3}\;\text{Tr}\left[(M_N-\slashed{K})(M_N-(\slashed{K}-\slashed{P})) \right]\nonumber\\
    &\times&\Tilde{\Delta}(K)\Tilde{\Delta}(K-P),
    \label{oneloopsigma}
\end{eqnarray}
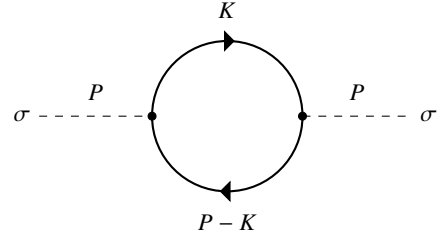
\begin{figure}[t]
    \centering
\begin{tikzpicture}
\draw[dashed] (-0.5,0) -- (1,0) node[midway,above=0.1cm] {$P$};
\draw[dashed] (3,0) -- (4.45,0) node[midway,above=0.1cm] {$P$};
\draw[-triangle 90] (2,1) -- +(.1,0);
\draw[-triangle 90] (2,-1) -- +(-.1,0);
\draw[black,thick] (2,0) circle (1);
\filldraw [black] (1,0) circle (1.5pt);
\filldraw [black] (3,0) circle (1.5pt);
\node [left] at (-0.5,0) {$\sigma$};
\node [right] at (4.45,0) {$\sigma$};
\node [above] at (2,1.2) {$K$};
\node [below] at (2,-1.2) {$P-K$};
\end{tikzpicture}
\caption{One-loop self energy of $\sigma$-meson in the linear RMF model.}
\label{sigma_Selfenergy}
\end{figure}
where $\Tilde{\Delta}(K)$ is the Matsubara fermion propagator and we have defined
\begin{eqnarray}
       P&=(i\omega,\Vec{p}),\;\;K&=(i\Tilde{\omega}_n+\mu,\Vec{k}), 
\end{eqnarray}
with $\Tilde{\omega}_n$ a fermion Matsubara frequency. The loop is made up of a nucleon-antinucleon pair, whose vacuum mass {is} grater than the temperature. Under these circumstances, the usual Hard Thermal Loop (HTL) approximation can not be implemented, and to obtain analytical results, a different approximation scheme is needed.

Direct computation {of} the trace in Eq.~(\ref{oneloopsigma}) yields
\begin{eqnarray}
    \Pi&=&-4g_{\sigma}^2T\sum_n\int\frac{d^3k}{(2\pi)^3}\;\left[M_N^2+\omega_n^2+k^2-\omega_n\omega-\Vec{k}\cdot\Vec{p} \right]\nonumber\\
    &\times&\Tilde{\Delta}(K)\Tilde{\Delta}(K-P),  \end{eqnarray}
which can be separated into three different contributions
    \begin{eqnarray}
    \Pi_1&\!\!=\!\!&-4g_{\sigma}^2T\sum_n\int\frac{d^3k}{(2\pi)^3}\left[M_N^2+\omega_n^2+k^2\right]\Tilde{\Delta}(K)\Tilde{\Delta}(K-P)\nonumber\\
    \Pi_2&\!\!=\!\!&4g_{\sigma}^2T\sum_n\int\frac{d^3k}{(2\pi)^3}\left[\Vec{k}\cdot\Vec{p} \right]\Tilde{\Delta}(K)\Tilde{\Delta}(K-P)\nonumber\\
    \Pi_3 &\!\!=\!\!&4g_{\sigma}^2T\sum_n\int\frac{d^3k}{(2\pi)^3}\left[\omega_n\omega \right]\Tilde{\Delta}(K)\Tilde{\Delta}(K-P). \end{eqnarray}
Each of these terms represents a different power of the Matsubara frequency that needs to be summed over and can be calculated by the standard methods described for example in Ref.~\cite{Bellac:2011kqa}. The $\Pi_1$ term can be simplified by noting that
\begin{eqnarray}
    \Pi_1&=&-g_{\sigma}^24T\sum_n\int\frac{d^3k}{(2\pi)^3}\;\Tilde{\Delta}(K)^{-1}\Tilde{\Delta}(K)\Tilde{\Delta}(K-P)\nonumber\\
    &=&-4g_{\sigma}^2T\sum_n\int\frac{d^3k}{(2\pi)^3}\;\Tilde{\Delta}(K-P).
\end{eqnarray}
Then, performing the sum, we obtain
\begin{eqnarray}
\Pi_1&=&-2g_\sigma^2\int\frac{d^3k}{(2\pi)^3}\left[1-\Tilde{n}_+(E_1)-\Tilde{n}_-(E_1) \right],
\end{eqnarray}
where we have defined $E_1=\sqrt{M_N^2+k^2}$ and  the Fermi-Dirac distribution
\begin{eqnarray}
    \Tilde{n}_\pm(E)=\frac{1}{e^{(E\mp\mu)/{T}}+1}.
\end{eqnarray}

For the other two contributions, $\Pi_2$ and $\Pi_3$, we obtain
\begin{eqnarray}
    \Pi_2=&=&4g_{\sigma}^2\int\frac{d^3k}{(2\pi)^3}\frac{\: \Vec{p}\cdot\Vec{k}}{E_1E_2}\nonumber\\
&\times&\Bigg{\{}\left[ \frac{1-\Tilde{n}_+(E_1)-\Tilde{n}_-(E_2)}{i\omega-E_1-E_2} -  \frac{1-\Tilde{n}_-(E_1)-\Tilde{n}_+(E_2)}{i\omega+E_1+E_2} \right]\nonumber\\
       &-&\left[ \frac{\Tilde{n}_-(E_1)-\Tilde{n}_-(E_2)}{i\omega+E_1-E_2} -  \frac{\Tilde{n}_+(E_1)-\Tilde{n}_+(E_2)}{i\omega-E_1+E_2} \right]\Bigg{\}},
       \label{Pi2_1}
\end{eqnarray}

\begin{eqnarray}
    \Pi_3=&=&4g_{\sigma}^2\int\frac{d^3k}{(2\pi)^3}\frac{i\omega}{E_2}\nonumber\\
&\times&\Bigg{\{}\left[ \frac{1-\Tilde{n}_+(E_1)-\Tilde{n}_-(E_2)}{i\omega-E_1-E_2} +  \frac{1-\Tilde{n}_-(E_1)-\Tilde{n}_+(E_2)}{i\omega+E_1+E_2} \right]\nonumber\\
       &-&\left[ \frac{\Tilde{n}_-(E1)-\Tilde{n}_-(E2)}{i\omega+E_1-E_2} +  \frac{\Tilde{n}_+(E1)-\Tilde{n}_+(E2)}{i\omega-E_1+E_2} \right]\Bigg{\}},
       \label{Pi3_1}
\end{eqnarray}
where $E_2=\sqrt{M_N^2+k^2+p^2-\Vec{p}\cdot\Vec{k}}$. 

To determine the correct approximation scheme for the present case where the $M_N>T$ we first define 
\begin{eqnarray}
\xi^2&=&k^2+M_N^2\nonumber\\
{k}dk&=&\xi d\xi
\end{eqnarray}

Notice that $\xi$ is always larger than $T$. Therefore, in a manner similar to the HTL approximation, where the loop momentum is considered to be of the order of $T$ and the external momentum is small compared to this scale, here we {consider} that the integration variable $\xi$ is larger than any of the external momentum components. Therefore, we can write the energy of the nucleon anti-nucleon in the loop as 
\begin{eqnarray}
  E_1&=&\sqrt{M_N^2+k^2}=\xi\nonumber\\
  E_2&=&\sqrt{M_N^2+k^2+p^2-2pk\cos\theta}\nonumber\\
  &=&\sqrt{\xi^2-2\sqrt{\xi^2-M_N^2}p\cos\theta+p^2}\nonumber\\ 
            &\approx&\xi-p\cos\theta+\frac{p^2-p^2 {\cos^2\theta}}{2 \xi }.
            \label{aprox1}
\end{eqnarray}
On the other hand, since $\xi > M_N$, we can approximate the Fermi-Dirac distributions as
\begin{eqnarray}
\tilde{n}_\pm(E_1)&\approx& e^{-\frac{\xi}{T}\pm\frac{\mu}{T}}\equiv n_{1\pm}\nonumber\\
\tilde{n}_\pm(E_2)&\approx&
n_{1\pm} -p\cos\theta \frac{dn_{1\pm}}{d\xi}\equiv n_{2\pm} .
\label{aprox2}
\end{eqnarray}
These approximations are analogous to the HTL approximation. Therefore, we can integrate over $\xi$ following a procedure similar to the case of the HTL approximation.

The temperature-dependent part of the $\Pi_1$ term can be easily computed. Using Eqs.~(\ref{aprox1}) and~(\ref{aprox2}), we obtain
\begin{eqnarray}
    \Pi_1&=&2 g_\sigma^2\int\frac{d\xi d(\cos\theta)}{(2\pi)^2} \sqrt{\xi ^2-M_N^2}  \cosh \left(\frac{\mu }{T}\right)\nonumber\\
    &=&\frac{2 g_\sigma^2}{\pi^2} M_N T\; K_1\left(\frac{M_N}{T}\right) \cosh \left(\frac{\mu }{T}\right),
\end{eqnarray}
where $K_1(x)$ is a modified Bessel function of {the} second kind.


The calculations of $\Pi_2$ and $\Pi_3$ are more involved. The process can be simplified defining the auxiliary functions
\begin{equation}
        \begin{aligned}
            \text{I}_1&=\int d\xi\;(\xi^2-M_N^2)\xi\;\frac{(n_{1+} +n_{2-})}{E_1\;E_2(i\omega-\Delta E^+)},\\
             \text{I}_2&=\int d\xi\;(\xi^2-M_N^2)\xi\;\frac{(n_{1-}+n_{2+})}{E_1\;E_2(i\omega+\Delta E^+)},\\
              \text{I}_3&=\int d\xi\;(\xi^2-M_N^2)\xi\;\frac{(n_{1-}-n_{2-})}{E_1\;E_2(i\omega+\Delta E^-)},\\
               \text{I}_4&=\int d\xi\;(\xi^2-M_N^2)\xi\;\frac{(n_{1+}-n_{2+})}{E_1\;E_2(i\omega-\Delta E^-)},\\
             \end{aligned}
             \label{IsFun}
    \end{equation}
and
    \begin{equation}
        \begin{aligned}
            \text{J}_1&=\int d\xi\;\sqrt{\xi^2-M_N^2}\;\xi\;\frac{(n_{1+} +n_{2-})}{E_2(i\omega-\Delta E^+)},\\
             \text{J}_2&=\int d\xi\;\sqrt{\xi^2-M_N^2}\;\xi\;\frac{(n_{1-}+n_{2+})}{E_2(i\omega+\Delta E^+)},\\
              \text{J}_3&=\int d\xi\;\sqrt{\xi^2-M_N^2}\;\xi\;\frac{(n_{1-}-n_{2-})}{E_2(i\omega+\Delta E^-)},\\
               \text{J}_4&=\int d\xi\;\sqrt{\xi^2-M_N^2}\;\xi\;\frac{(n_{1+}-n_{2+})}{E_2(i\omega-\Delta E^-)},\\
             \end{aligned}
             \label{JsFun}
    \end{equation}
where 
\begin{equation}
    \begin{aligned}
        \Delta E^+ &= E_1+E_2,\\
        \Delta E^- &= E_1-E_2.
    \end{aligned}
\end{equation}
Then we can write 
\begin{eqnarray}
    \Pi_2 =\frac{4g_\sigma^2 p}{(2\pi)^2}\int d(\cos\theta)\cos\theta \left[(\text{I}_1-\text{I}_2) - (\text{I}_3-\text{I}_4)\right],
   \end{eqnarray}
   \begin{eqnarray}
    \Pi_3 =i\omega\frac{4g_\sigma^2 }{(2\pi)^2}\int d(\cos\theta)\left[(\text{J}_1+\text{J}_2) + (\text{J}_3+\text{J}_4)\right].
\end{eqnarray}

We now proceed to the explicit calculation of $\Pi_2$ and $\Pi_3$.

\subsection{Calculation of $\Pi_2$}

 Defining the variable $y\equiv \cos\theta$, the explicit form of $I_3$ is
 
\begin{eqnarray}
   I_3&=&\frac{4p}{T}\int d\xi    y^2 \left(\xi ^2-M_N^2\right) e^{-\frac{\mu +\xi }{T}}\nonumber\\
  &\times& \frac{\xi^2}{ 2 \xi ^2-p^2 \left(y^2-1\right)-2 \xi  p y },\nonumber\\
  &\times&\frac{1}{p^2 \left(y^2-1\right)+2\xi(p y+i\omega)}
\end{eqnarray}

which can be decomposed into partial fractions, giving
\begin{eqnarray}
    I_3&=&\frac{4p}{T}\int d\xi  y^2 \left(\xi ^2-M_N^2\right) e^{-\frac{\mu +\xi }{T}}\nonumber\\
   &\times& \Bigg{[}\frac{\alpha}{p^2 \left(y^2-1\right)+2\xi  ( p y+ i\omega)}\nonumber\\
  &+& \delta\frac{\alpha+\beta\xi}{\xi-z_1} -\delta\frac{\alpha+\beta\xi}{\xi-z_2}\Bigg{]},
  \label{I3ss}
\end{eqnarray}
where
\begin{eqnarray}
    \alpha&\equiv&\frac{p^2 \left(y^2-1\right)}{4 p i\omega y+4 (i\omega)^2-2 p^2 \left(y^2-1\right)},\nonumber\\
    \beta&\equiv& {\frac{i\omega}{p \left(p y^2-p-2 i\omega y\right)-2 (i\omega)^2}},\nonumber\\
    \delta&\equiv&\frac{1}{\sqrt{p^2 \left(3 y^2-2\right)}},
\end{eqnarray}
and where $z_1$ and $z_2$, given by
\begin{eqnarray}
    z_1&=&\frac{1}{2} \left(p y-\sqrt{3 p^2 y^2-2 p^2}\right),\nonumber\\
    z_2&=&\frac{1}{2} \left(py+\sqrt{3 p^2 y^2-2 p^2}\right),
\end{eqnarray}
are the roots of the polynomial $2 \xi ^2-p^2 \left(y^2-1\right)-2 \xi  p y$. {The leading order result,} for the case when $M_N\gg T$, {is given by}  
\begin{eqnarray}
    I_3&\!\!=\!\!& \frac{p^3 y^2 \left(y^2-1\right) e^{\frac{\mu -M_N}{T}} \left(p^2 \left(y^2-1\right)-2 p T y-2 (i\omega) T\right)}{2 (p y+i\omega)^2 \left(-p^2 \left(y^2-1\right)+2 p (i\omega) y+2 (i\omega)^2\right)} \nonumber\\
    &\!\!+\!\!& \frac{2 p y^2 e^{-\frac{\mu +{M_N}}{T}} \left(p^2 \left(y^2-1\right)-2 p (i\omega) y-2 (i\omega) T\right)}{p^2 \left(y^2-1\right)-2 p (i\omega) y-2 (i\omega)^2}.
\end{eqnarray}

Within this strategy, $I_1$ and $I_2$ turn out to be proportional to $1/M_N$, and thus, they can be neglected, while $I_4$ can be computed {in a similar way.} Integrating over $y$ and following the partial fraction method, $\Pi_2$ can be approximated as   
 \begin{eqnarray}
\Pi_2&=&(i\omega)\frac{2 g_\sigma^2 e^{-\frac{M_N}{T}}}{\pi^2} \Bigg{(}\frac{\left(48 x^4+26 x^2+1\right) }{3 x^2 \left(x^2+1\right)}\nonumber\\
    &+&\frac{ \left(22 x^4+13 x^2+1\right) \log \left(\frac{x+1}{x-1}\right)}{2 \left(x^3+x\right)}\nonumber\\
    &+&\frac{ \left(2 x^2+1\right) \left(19 x^2+5\right)  \log \left(\frac{ \sqrt{1+3x^2}+x-1}{ \sqrt{ 1+3x^2}-x+1}\right)}{ \left(x^2+1\right) \sqrt{ \left(1+3x^2\right)}}\Bigg{)}\nonumber\\
    &\times&\left(i\omega \cosh \left(\frac{\mu }{T}\right)-T \sinh \left(\frac{\mu }{T}\right)\right),
\end{eqnarray}
where we have defined $x\equiv\frac{i \omega}{p}$.

\subsection{Calculation of $\Pi_3$}

As we did for the case of Eq.~(\ref{I3ss}), we can decompose $J_3$ in partial fractions
\begin{eqnarray}
    J_3&=&\frac{4p}{T}\int d\xi  \xi   y \sqrt{\xi ^2-M_N^2} e^{-\frac{\mu +\xi }{T}}\nonumber\\
    &\times& \Bigg{[}\frac{\alpha}{p^2 \left(y^2-1\right)+2\xi  ( p y+ i\omega)}\nonumber\\
  &+& \delta\frac{\alpha+\beta\xi}{\xi-z_1} -\delta\frac{\alpha+\beta\xi}{\xi-z_2}\Bigg{]}.
\end{eqnarray}

Notice that he smallest value that $\xi$ takes is $M_N$, thus, we obtain
\begin{eqnarray}
    1\leq\frac{p^2 \left(y^2-1\right)}{2\xi  ( p y+ i\omega)}\leq1,\ \ \ \ \ \
    1\leq&\frac{z_i}{\xi}\leq 1,
\end{eqnarray}
with $i=1,2$. This {allows} us to write
\begin{eqnarray}
     J_3&=& \frac{4p}{T}\int d\xi y \sqrt{\xi ^2-M_N^2} e^{-\frac{\mu +\xi }{T}}\nonumber\\
     &\times&\Bigg{[}\frac{1}{2  ( p y+ i\omega)}\frac{\alpha}{\frac{p^2 \left(y^2-1\right)}{2\xi  ( p y+ i\omega)}+1}\nonumber\\
  &+& \delta\frac{\alpha+\beta\xi}{1-\frac{z_1}{\xi}} -\delta\frac{\alpha+\beta\xi}{1-\frac{z_2}{\xi}}\Bigg{]}\nonumber\\
  &=&\frac{4p}{T}\int d\xi y \sqrt{\xi ^2-M_N^2} e^{-\frac{\mu +\xi }{T}}\nonumber\\
  &\times&\Bigg{[}\frac{1}{2  ( p y+ i\omega)}\alpha\sum_{n=0}^\infty \left(\frac{p^2 \left(y^2-1\right)}{2\xi  ( p y+ i\omega)}\right)^n\nonumber\\
  &+& \delta(\alpha+\beta\xi)\left(\sum_{n=0}^\infty\left(-\frac{z_1}{\xi}\right)^n -\sum_{n=0}^\infty\left(-\frac{z_2}{\xi}\right)^n \right)\Bigg{]}.
\end{eqnarray}

At leading order, we obtain
\begin{eqnarray}
    J_3 &\!\!\approx\!\!&\frac{4p}{T}\int d\xi y \sqrt{\xi ^2-M_N^2} e^{-\frac{\mu +\xi }{T}}\frac{\alpha}{2  ( p y+ i\omega)}\nonumber\\
    &\!\!=\!\!&\frac{p^3 x \left(x^2-1\right) e^{-\frac{\mu }{T}} K_1\left(\frac{M_N}{T}\right)}{(p x+i\omega) \left(-\left(p^2 \left(x^2-1\right)\right)+2 p i\omega x+2 (i\omega)^2\right)}.
\end{eqnarray}
In analogy to the calculation of $\Pi_2$,  $J_1$ and $J_2$ are proportional to $1/M_N$ therefore, they can also be neglected. Calculating $J_4$ in the same way and integrating over $y$, we can express $\Pi_3$ as
\begin{eqnarray}
    \Pi_3&=&\frac{2 g_\sigma^2 M_N {(i\omega)} K_1\left(\frac{M_N}{T}\right)}{3\pi^2} \sinh \left(\frac{\mu }{T}\right) \nonumber\\
    &\times&\Bigg{[}30 x^2-4+3 \left(5 x^3+x\right) \log \left(\frac{x+1}{x-1}\right) \nonumber\\
    &-&\frac{6 x^2 \left(5 x^2+1\right) \log \left(\frac{x^2+\sqrt{3 x^2+1}+1}{x^2-\sqrt{3 x^2+1}+1}\right) }{ \sqrt{3 x^2+1}}\Bigg{]},
\end{eqnarray}
where once again  $x\equiv\frac{i \omega}{p}$.

\subsection{Final expression for the $\sigma$ self-energy}

Comparing $\Pi_1$ and $\Pi_3$ with $\Pi_2$, we notice that $\Pi_2$ is of the order of zero in $M_N$ while $\Pi_1$ and $\Pi_3$ are of the first order. Therefore, we can safely disregard $\Pi_2$ compared to $\Pi_1$ and $\Pi_3$. With this last consideration, the $\sigma$ self-energy can be written as
\begin{eqnarray}
    \Pi&=&M_T^2+\gamma_T(i\omega)\Bigg{\{}10x^2-\frac43 \nonumber\\
    &+& \left(5 x^2+1\right)\Bigg{[} x\log \left(\frac{x+1}{x-1}\right) \nonumber\\
    &-&\frac{2 x^2 }{ \sqrt{3 x^2+1}}\log \left(\frac{x^2+\sqrt{3 x^2+1}+1}{x^2-\sqrt{3 x^2+1}+1}\right)\Bigg{]}\Bigg{\}}.
    \label{sigmaself}
\end{eqnarray}
where we have introduced the definitions
\begin{eqnarray}
    M_T^2&\equiv&\frac{2 g_\sigma^2 M_N T}{\pi^2} K_1\left(\frac{M_N}{T}\right) \cosh \left(\frac{\mu }{T}\right),\nonumber\\
    \gamma_T&\equiv&  
    \frac{M_T^2 \tanh{\left(\frac{\mu }{T}\right)}}{T}.
\end{eqnarray}

\begin{figure}[t]
        \centering
\includegraphics[width=1\linewidth]{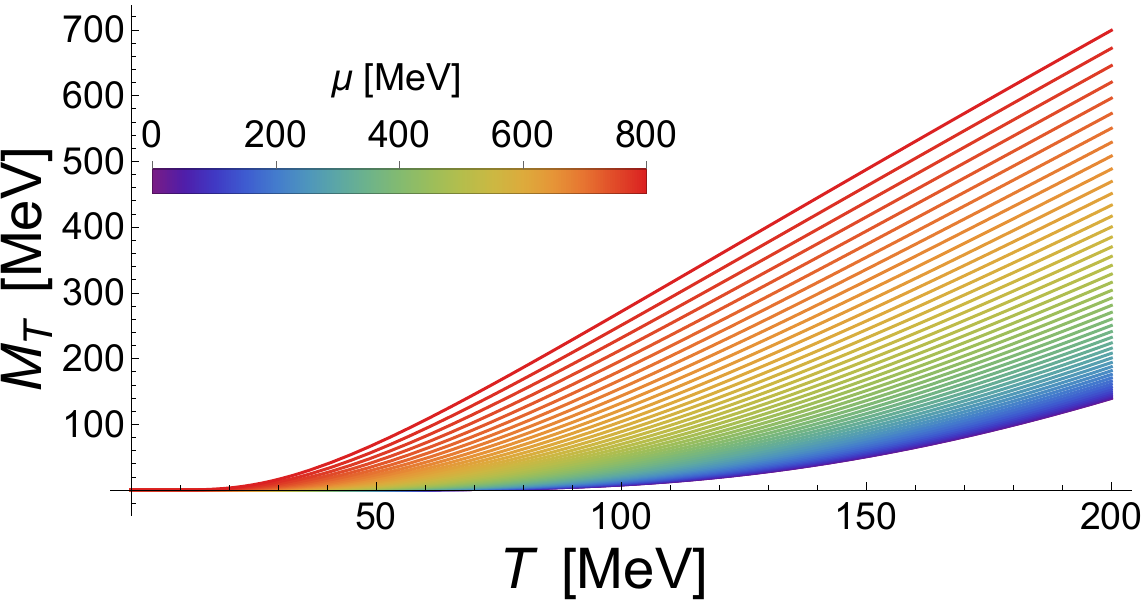}
\caption{$M_T$ as a function of temperature for different values of the baryon chemical potential $\mu$. We take $g_\sigma=10.5$ and $m_\sigma=510$ MeV as an average of the estimations {reported in Refs.}~\cite{Long:2003dn,Sugahara:1993wz}.}
\label{MT}
\end{figure}

\begin{figure}[t]
        \centering
\includegraphics[width=1\linewidth]{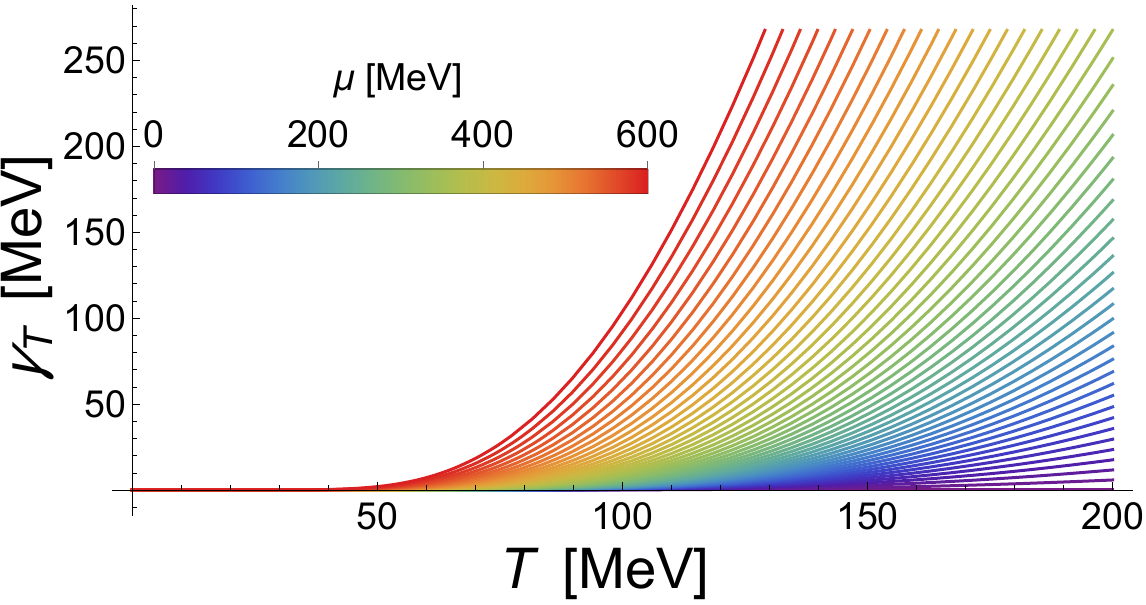}
\caption{$\gamma_T$ as a function of temperature for different values of the baryon chemical potential $\mu$. We take $g_\sigma=10.5$ and $m_\sigma=510$ MeV as an average of the estimations {reported in Refs.}~\cite{Long:2003dn,Sugahara:1993wz}.}
\label{gammaT}
\end{figure}

As we show below, $M_T$ plays the role of a thermal mass, whereas $\gamma_T$ is related to Landau damping. {Figures~\ref{MT} and~\ref{gammaT} show the behavior of $M_T$ and $\gamma_T$, respectively, as functions of $T$, for different values of $\mu$, for $m_\sigma=510$ MeV and $g=10.5$.} Using these results, the effective one-loop propagator of $\sigma$-mesons in a medium where $M_N\gg T$ is written as in Eq.~(\ref{effectiveprop}), 
with $\Pi$ given by Eq.~(\ref{sigmaself}).

\section{Analytic properties}\label{III}

To study the properties of the propagator, we perform the analytic continuation back to Minkowski space $i\omega\rightarrow p_0$, for which $x\to p_0/p$ with $p_0$ real. Notice that for $-1<x<1$
\begin{eqnarray}
\frac{x+1}{x-1}<0,\ \ \ \ \ \
\frac{x^2+\sqrt{3 x^2+1}+1}{x^2-\sqrt{3 x^2+1}+1}<0,
\end{eqnarray}
and thus the logarithmic terms develop an imaginary part, explicitly 
   \begin{eqnarray}
       \log \left(\frac{x+1}{x-1}\right)=\log \Bigg{|}\frac{x+1}{x-1}\Bigg{|}
       +i \pi\theta\left(1-x^2 \right),
   \end{eqnarray}
and
   \begin{eqnarray}
       \log \left(\frac{x^2+\sqrt{3 x^2+1}+1}{x^2-\sqrt{3 x^2+1}+1}\right)&=&\log \Bigg{|}\frac{x^2+\sqrt{3 x^2+1}+1}{x^2-\sqrt{3 x^2+1}+1}\Bigg{|}\nonumber\\
       &+&i \pi\theta\left(1-x^2 \right),
   \end{eqnarray}  
where $\theta$ is the Heaviside function. 
Therefore, the propagator can be expressed as
\begin{eqnarray}
\Delta^*(p_0,p)=\frac{-1}{P^2-M_\sigma^2-\gamma_Tp_0F(x)-i\pi A(x)\theta(p^2-p_0^2)},\nonumber\\
\label{Propagator}
\end{eqnarray}
where we have defined
\begin{eqnarray}
    P^2=p_0^2-p^2,\ \ \ \ M_\sigma^2=m_\sigma^2+M_T^2
   \end{eqnarray}
    \begin{eqnarray}     
    F(x)&=&10x^2-\frac43 + \left(5 x^2+1\right)\Bigg{[} x\log \Bigg{|}\frac{x+1}{x-1}\Bigg{|} \nonumber\\
    &-&\frac{2 x^2 }{ \sqrt{3 x^2+1}}\log \Bigg{|}\frac{x^2+\sqrt{3 x^2+1}+1}{x^2-\sqrt{3 x^2+1}+1}\Bigg{|}\;\Bigg{]},
    \end{eqnarray}
    and
   \begin{eqnarray}
        A(x)
        &=&\left(5 x^2+1\right) \left(x-\frac{2 x^2  }{ \sqrt{3 x^2+1}}\right).
    \end{eqnarray}
Written in this form, we can easily identify the real and imaginary parts of the propagator, from which we can find the dispersion relation $\omega(p)$ and the damping rate $\gamma(p)$ as
\begin{eqnarray}
\omega^2(p)&=&p^2+m_\sigma^2+\text{Re}\left[\Pi\Big{|}_{p_0=\omega(p)}\right]\nonumber\\
&=&p^2+m_\sigma^2+M_T^2+\gamma_T\omega(p)F\left(\omega(p)/p\right),
\label{DispersionLaw}
\end{eqnarray}
\begin{eqnarray}
    \gamma(p)&=&-\frac{1}{2\omega(p)}\text{Im}\left[\Pi\Big{|}_{p_0=\omega(p)}\right]\nonumber\\
    &=&\frac{\gamma_T}{2}\pi A\left(\omega(p)/p\right)\theta(p^2-\omega(p)^2).
    \label{DampingRate}
\end{eqnarray}

It is possible to obtain approximate analytical solutions for the dispersion relation for small and large values of $p$. First, let us rewrite Eq.~(\ref{DispersionLaw}) as
\begin{eqnarray} \frac{\omega(q)^2}{p^2}-1-\frac{M_\sigma^2}{p^2}-\frac{\gamma_T}{p}\frac{\omega(p)}{p}F(\omega(p)/p)&=&0\nonumber\\
 x^2-1-\frac{M_\sigma^2}{p^2}-\frac{\gamma_T}{p}xF(x)&=&0.
 \label{DispersionLaw2}
\end{eqnarray}
For small $p$, $\gamma_T/p>1$ and the function $F(x)$ must be small enough such that Eq.~(\ref{DispersionLaw}) has solutions. Hence, we have to look for the behavior of $F(x)$ when $x\rightarrow\infty$. At  leading order, one gets \begin{eqnarray}
\omega(p)=p^2+M_\sigma^2.
\end{eqnarray}    
Notice that in the limit $p\rightarrow0$, $F(p_0/p)\rightarrow0$, therefore,
\begin{eqnarray}
\omega^2(p= 0)=m_\sigma^2+M_T^2,
\label{approxSmall}
\end{eqnarray}
and thus the square of the $\sigma$ vacuum mass increases by $M_T^2$ which justifies the name \lq\lq thermal mass" for $M_T$. 
\begin{figure}[t]
\centering
\includegraphics[width=1\linewidth]{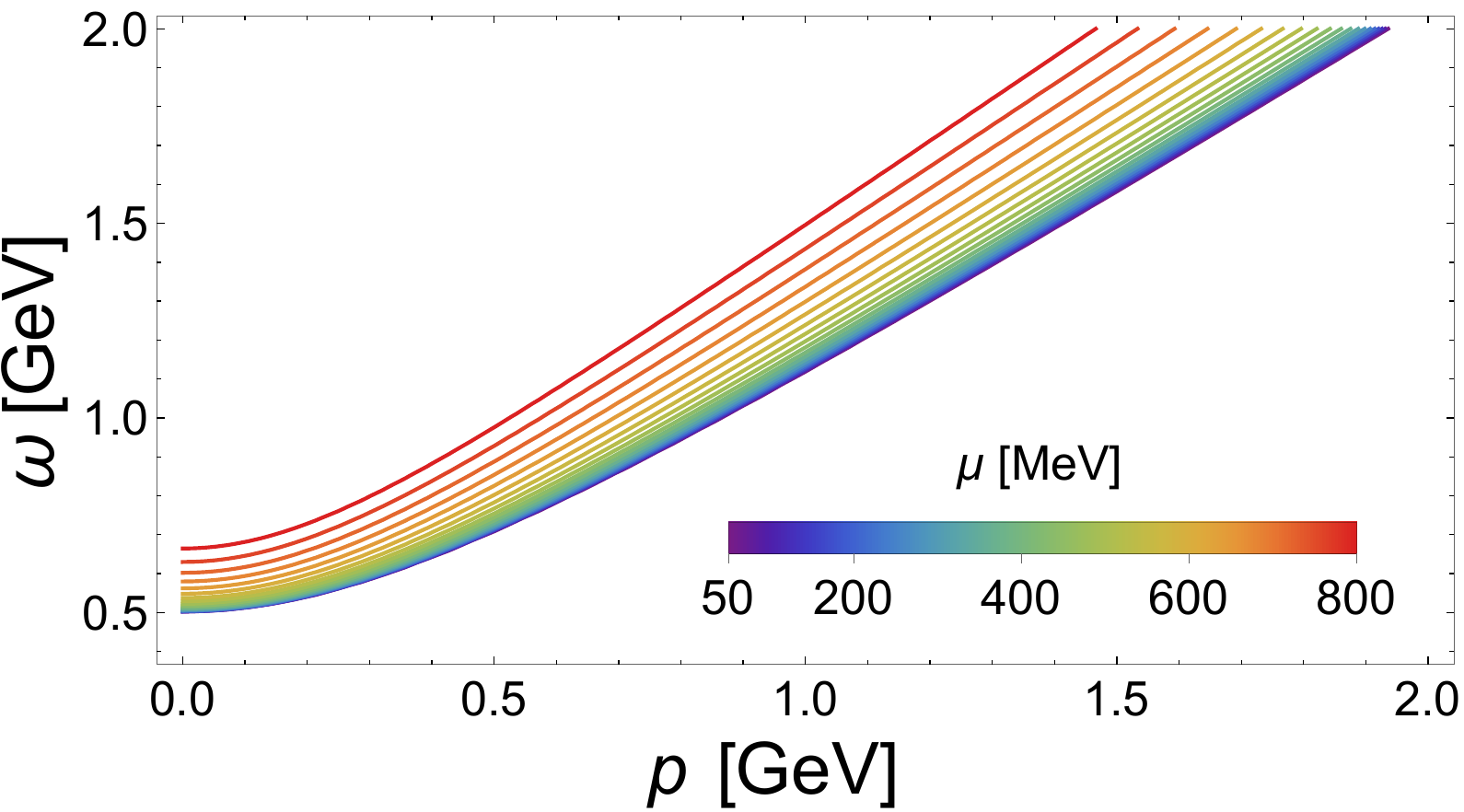}
\caption{Numerical solution to Eq.~(\ref{DispersionLaw}) using  $T=$150 MeV, $g_\sigma=10.5$ and $m_\sigma=510$ MeV as an average of the estimations done in~\cite{Long:2003dn,Sugahara:1993wz} for different values of $\mu$. Notice that the $\sigma$ energy increases as $\mu$ increases.}
\label{Fig:NumericSolution1}
\end{figure}
For large $p$ the procedure is analogous. In this case $\gamma_T/p<1$, and in this limit, to {find} solutions, the function $F(x)$ must be large enough. Hence, we look for the behavior of $F(x)$ for $x\rightarrow1$. The leading order for this approximation is given by
\begin{eqnarray}
\omega^2(p)=p^2+M_\sigma^2- \alpha_{n}\gamma_T \omega(p),
\label{approxGreat}
\end{eqnarray}       
where $\alpha_n$ is the constant $\alpha_n=\frac{2}{3} (13-18 \log (2))\approx$ 0.3489. The solution {to}  Eq.~(\ref{approxGreat}) is 
\begin{eqnarray}
    \omega(p)=\frac{\alpha_n\gamma_T}{2}\pm\sqrt{\frac{\alpha_n^2 \gamma_T^2}{4}+ M_\sigma^2+ p^2}.
    \label{solutions}
\end{eqnarray}
The solution for positive energies is obtained from Eq.~(\ref{solutions}) and corresponds to the positive sign of the square root
\begin{eqnarray}
    \omega_+(p)=\frac{\alpha_n\gamma_T}{2}+\sqrt{\frac{\alpha_n^2 \gamma_T^2}{4}+ M_\sigma^2+ p^2}.
    \label{solplus}
\end{eqnarray}
The solution for negative energies is obtained using the negative sign for the square root in Eq.~(\ref{solutions}), with the change $\mu\rightarrow-\mu$, which changes $\gamma_T\rightarrow-\gamma_T$. The explicit solution of Eq.~(\ref{approxGreat}) for negative energies is thus
\begin{eqnarray}
    \omega_-(p)=-\frac{\alpha_n\gamma_T}{2}-\sqrt{\frac{\alpha_n^2 \gamma_T^2}{4}+ M_\sigma^2+ p^2}
    \label{solminus},
\end{eqnarray}
which shows that $\omega_+(p)=-\omega_-(p)$. This last relation is valid in general for the {red}{whole} $p$-domain.
The behavior of the numerical solution for the positive energy dispersion relation, as a function of $p$, for several values of $\mu$ and a fixed temperature $T=150$ MeV, using $m_\sigma=510$ MeV and $g=10.5$, is shown in Fig.~\ref{Fig:NumericSolution1}.

\subsection{Spectral density}

We now explore the expression for the spectral density,
\begin{equation}
\rho(p_0,p)=2 \text{Im}\Delta^*(q_0+i\eta,q).
\end{equation}
Following the standard methods described in Ref.~\cite{Bellac:2011kqa}, it is easy to show that
\begin{eqnarray}
\rho(p_0,p)&=&2\pi Z(\omega(p))\left[ \delta(p_0 - \omega(p)) - \delta(p_0+\omega(p)) \right] \nonumber\\
&+&\beta(p_0,p),
\end{eqnarray}
where $Z(\omega(p))$ is the residue 
of the propagator at the poles $\omega(p)$ and $\beta$ is the Landau damping term
\begin{eqnarray}
\beta(p_0,p)&\!\!\!=\!\!\!&2\pi\gamma_T p_0 A(x)\theta\left(1-x^2\right)\nonumber\\
&\!\!\!\times\!\!\!&\left[\left(P^2\!-\!M_\sigma^2\!-\!\gamma_T p_0 F\left(x\right)\right)^2\!\!-\Big{(}\gamma_T p_0A\left(x\right)\pi\Big{)}^2\right]^{-1}\!\!\! .\nonumber\\
\end{eqnarray}
Once again, we can provide explicit analytic approximations for the residues in the small and large $p$-domains. For small $p$ the residue behaves as
\begin{eqnarray}
    Z(p)\approx\frac{1}{2\sqrt{p^2+M_\sigma^2}}.
\end{eqnarray}
On the other hand, for the large $p$-domain, the approximation is
\begin{eqnarray}
    Z(p)\approx\frac{1}{2\sqrt{p^2+M_\sigma^2+\frac{\alpha_n^2 \gamma_T}{4}}}.
\end{eqnarray}

\begin{figure}[t]
    \centering
    \begin{tikzpicture}
        \draw [-{Triangle}, thick] (0,0) -- (2.75,0);
        \draw [-{Triangle}, thick] (0,0) -- (-2.75,0);
        \draw [-{Triangle}, thick] (0,0) -- (0,2.75);
         \draw [-{Triangle}, thick] (0,0) -- (0,-2.75);
         \node [above] at (3.3,-0.2) {Re[$p_0$]};
         \node [above] at (0,2.75) {Im[$p_0$]};
      
       \draw [blue, ultra thick] (2.2,0.4) arc (10.4757:349.524:2.2);
       \draw [blue, ultra thick] (-1.6,0.4) -- (2.2,0.4);
        \draw [blue, ultra thick] (-1.6,-0.4) -- (2.2,-0.4);
        \draw [blue, ultra thick] (-1.6,-0.4) -- (-1.6,0.4);
        \draw [blue, ultra thick,->] (2.2,0.4) arc (10.4757:60:2.2);
        \filldraw [red] (1,0) circle (2.5pt);
        \filldraw [red] (-1,0) circle (2.5pt); 
        \node [above] at (1.2,2) {$\Gamma$};
        \node [above] at (1,-0.45) {$p_0=\omega_+$};
        \node [above] at (-1,-0.45) {$p_0=\omega_-$};
         \node [above] at (2.1,0) {$\delta$};
        \draw [thick] (2,0)--(2,0.4) ;
    \end{tikzpicture}
    \caption{Integration contour in the complex $p_0$-plane used in Eqs.~\eqref{eq:Contour1} and~(\ref{eq:Sum0}).}
 \label{fig:Int_cont}
\end{figure}
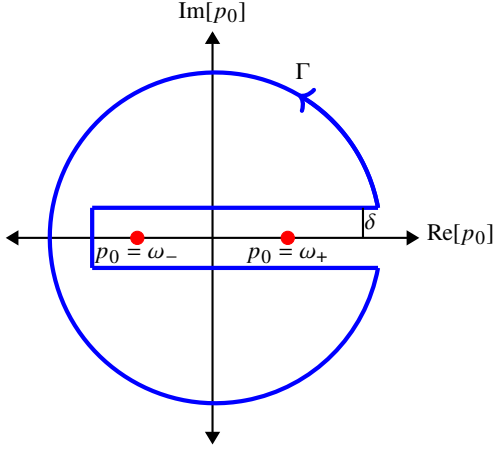

\subsection{Sum rules}

We can make use of the analytical properties to derive the sum rules obeyed by the propagator. Equation~(\ref{Propagator}) is analytic in the complex $p_0$-plane cut from $-p$ to $p$. Moreover, it has simple poles at $p_0=\omega(p)$. We can use Cauchy's theorem to write Eq.~(\ref{Propagator}) as 
  \begin{eqnarray}
         \Delta^*(p_0,p)&=&\oint_\Gamma\frac{d z}{2i\pi}\frac{\Delta^*(z,p)}{z-p_0}\nonumber\\
         &=&\int_{-\infty}^{\infty}\frac{d p_0'}{2i\pi}\frac{\Delta^*(p_0+i\delta,p)-\Delta^*(p_0'-i\delta,p)}{p_0'-p_0}\nonumber\\
        &+& \oint_{\Gamma'}\frac{d z}{2i\pi}\frac{\Delta^*(z,p)}{z-p_0},
        \label{eq:Contour1}
     \end{eqnarray}
where $\Gamma$ is the contour depicted in Fig.~\ref{fig:Int_cont} and $\Gamma'$ is the circle at infinity. Eq.~(\ref{eq:Contour1}) can be rewritten in terms of the spectral density $\rho(p_0,p)$, which contains both the discontinuities across the cuts and the residues at the poles,
\begin{eqnarray}
    \rho(p_0,p)=2 \text{Im}\Delta^*(p_0+i\delta,p).
\end{eqnarray}
Therefore, \begin{eqnarray}
\!\!\!\Delta^*(p_0,p)&\!\!=\!\!&\int_{-\infty}^{\infty}\frac{d p_0'}{2\pi}\frac{\rho(p_0',p)}{p_0'-p_0}+ \oint_{\Gamma'}\frac{d z}{2i\pi}\frac{\Delta^*(z,p)}{z-p_0}.
         \label{eq:Sum0}
     \end{eqnarray}
Since $\Delta^*(z,p)\sim 1/z^2$ for $z\rightarrow\infty$, there is no contribution from $\Gamma'$. On the other hand,
\begin{eqnarray}
\lim_{p_0\to0} \Delta^*(p_0,p)=\frac{1}{p^2+m_\sigma^2+M_T^2}.
\end{eqnarray}
As a consequence,
\begin{eqnarray}
\int_{-\infty}^{\infty}\frac{d p_0'}{2\pi}\frac{\rho(p_0',p)}{p_0^\prime}=\frac{1}{p^2+m_\sigma^2+M_T^2},
\label{eq:Sum1}
\end{eqnarray}
which is the first of the sum rules obeyed by the spectral density. Notice that Eq.~(\ref{eq:Sum1}) is the analog of the sum rule that the photon transverse mode obeys in the HTL approximation. Other sum rules can be obtained considering the asymptotic expansion of
\begin{eqnarray}
\frac{1}{p_0'-p_0}=  -\frac{1}{p_0}\sum_{n=0}^\infty\left(\frac{p_0'}{p_0}\right)^{2n+1},
\end{eqnarray}
which allows us to express Eq.~(\ref{eq:Sum0}) as
\begin{eqnarray}
\Delta^*(p_0,p)&=&\int_{-\infty}^{\infty}\frac{d p_0'}{2\pi}\frac{\rho(p_0',p)}{p_0'-p_0}\nonumber\\
&=&-\frac{1}{p_0}\sum_{n=0}^\infty\int_{-\infty}^{\infty}\frac{d p_0'}{2\pi}\left(\frac{p_0'}{p_0}\right)^{2n+1}\!\!\!\rho(p_0',p).
\label{sum}
\end{eqnarray}
Comparing the expansion of Eq.~(\ref{Propagator}) in powers of $p_0^{-1}$ with Eq.~(\ref{sum}), we find the sum rules for different values of $n$. For $n=0$,
\begin{eqnarray}
\int_{-\infty}^{\infty}\frac{d p_0'}{2\pi}p_0'\ \rho(p_0',p)=1.
\label{eq:SumN0}
\end{eqnarray}
For $n=1$,
\begin{eqnarray}
\int_{-\infty}^{\infty}\frac{d p_0'}{2\pi}p^{\prime 3}_0\ \rho(p_0',p)=p^2+M_\sigma^2.
\label{eq:SumN1}
\end{eqnarray}
For $n=2$,
\begin{eqnarray}
\int_{-\infty}^{\infty}\frac{d p_0'}{2\pi}p_0'^5\ \rho(p_0',p)=(p^2+M_\sigma^2)^2.
\label{eq:SumN2}
\end{eqnarray}
Notice once more that Eqs.~(\ref{eq:SumN0}) and~(\ref{eq:SumN1}) are analogous to the corresponding sum rules obtained for the photon transverse mode in the HTL approximation. {It is remarkable that sum rules with this simple structure, similar to those obtained in the HTL approximation, can be obtained in the present analysis, which shows that the approximation scheme hereby adopted captures the leading behavior of the $\sigma$-meson thermal and baryon density modifications dispersion properties within a nucleon dominated environment.}

\section{Conclusions and outlook}\label{IV}

Particles evolving in a medium inherit the medium-properties through interactions. An effective description of the interactions is achieved by introducing modifications to the propagation properties of the interaction mediator. This strategy has been successfully applied in finite-temperature calculations in different theories, most notably in QCD where the gluon propagator is modified by temperature and baryon density effects. In this work, we have followed a similar strategy finding the temperature and baryon density modifications to one of the mediators of the strong interaction at low energies: the $\sigma$-meson. We have found the effective one-loop $\sigma$-propagator considering its interactions within a medium consisting of nucleons. We have resorted to an approximation that accounts for the large nucleon mass compared to the rest of the energy scales. We have shown that the $\sigma$ dispersion relation and the damping rate receive {non-negligible} thermal and baryon density contributions. The $\sigma$ mass develops a thermal mass component, and the propagator develops an imaginary part that contains a piece associated with the branch cut corresponding to Landau damping. We have also studied the analytical properties of the propagator, providing approximations for the dispersion relation and residue at the pole in the small- and large-momentum regimes. {We have shown that the spectral density obeys simple, HTL-like, sum rules, which shows that the adopted approximation scheme captures the leading order behavior of the $\sigma$-meson propagator in a nucleon dominated environment.}

We plan to use this propagator when describing the interaction of $\Lambda$s and nucleons, where the latter take part of the vortical motion produced in the low-density region (the corona) in a peripheral heavy-ion collision. This strategy is similar to the one we have already used to describe the transfer of the vortical motion to the spin of $\Lambda$s in the high-density region (core) of the collision, where an effective gluon propagator captures the properties of a deconfined medium~\cite{Ayala:2023vgv}. {The extra needed ingredient} is provided by a propagator that captures the vortical motion of the nucleons of the medium. This work is currently being pursued and will be reported elsewhere.

\section*{Acknowledgements}

Support for this work was received in part by UNAM-PAPIIT grant number IG100322 and by Consejo Nacional de Humanidades, Ciencia y Tecnolog\'ia grant number CF-2023-G-433.

\bibliography{bibliosigma}

\end{document}